\documentstyle[pre,aps]{revtex}
\begin{document}
\preprint{Growth2-Vers.2.5}
\draft
\title{A Soluble Phase Field Model}

\author{Umberto Marini Bettolo Marconi}
\address{Dipartimento di Matematica e Fisica, Universit\`a di Camerino,
         Via Madonna delle Carceri,I-62032 , Camerino, Italy}
\address{Unit\`a INFM di Camerino and Sez. INFN Perugia}

\author{Andrea Crisanti and Giulia Iori}
\address{Dipartimento di Fisica, Universit\`a di Roma ''La Sapienza''
         P.le A.Moro 2, I-00185 , Roma, Italy}
\address{Unit\`a INFM di Roma I}

\date{\today}
\maketitle

\begin{abstract}
The kinetics of an initially undercooled solid-liquid melt is studied
by means of a generalized Phase Field model, which describes the dynamics
of an ordering non-conserved field $\phi$ (e.g. solid-liquid order
parameter) coupled to a conserved field (e.g. thermal field). 
After obtaining the rules governing the 
evolution process, by means of analytical arguments, we present
a discussion of the asymptotic time-dependent solutions.
 The full solutions of the exact self-consistent
equations for the model are also obtained
and compared with computer simulation
results. In addition, in order to check
the validity of the present model we confronted its predictions
against those of the standard
Phase field mode and
found reasonable agreement. 
 Interestingly, we find that the system relaxes towards
a mixed phase, depending on the average value of the conserved field,
i.e. on the initial condition.
Such a phase is characterized by large fluctuations of
the $\phi$ field.

\end{abstract}

 In the last few years considerable effort has been devoted to the study 
of systems far from equilibrium \cite{Bray}. 
Well known examples are provided by phase 
separating systems, initially prepared in a
state of equilibrium, and rendered unstable by  
modifying a control parameter
such as temperature, pressure,
or magnetic field.  To restore stability they evolve towards a different
equilibrium state determined by the
final value of the controlling fields. Such evolution
can be very slow and is often characterized by non uniform, 
complex structures both in space and time. 

 Two simple dynamical models, often called Model A 
\cite{Glauber} and model B \cite{Cahn}, have been
introduced in the literature in order to understand kinetic ordering
phenomena (see also \cite{Hohenberg,GSMS}).
 The model A
describes the growth process, when the order parameter is non conserved,
whereas the  model B is appropriate if
the order parameter is conserved.
In the first case,
the late stage growth is
driven by the tendency of the system to minimize the
energy cost due to the presence of interfaces between 
regions separating different phases. Thus, as the curvature decreases
the process slows down and the domain size, $L(t)$, grows in time 
according to the law $L(t)\sim t^{1/2}$.
In the conserved case, instead,
the approach to equilibrium is limited by the diffusion of the
the aggregating material, as larger domains can grow only at
the expenses of smaller ones. The average size $L(t)$ increases 
proportionally to $t^{1/z}$, where the dynamical exponent $z$
is $3$ for scalar order parameters and $4$ for vector order parameters.

A further model, known as Phase field model (PFM), 
is somehow intermediate between A and B and consists
of two fields coupled bilinearly: one field represents 
a non conserved ordering
parameter, with  type A dynamics, whereas the second is a temperature 
shift field subject to a diffusion equation supplemented by a source
term.
 The model can be cast in the form of coupled partial differential
equations for a non conserved order parameter interacting with
a time dependent conserved field. Its dynamics is very rich, since
it displays features characterizing both the pure A and 
the pure B models as it is revealed 
from the analysis of the structure functions 
at different times. In other words, after a rapid initial 
evolution one observes an intermediate stage in which the growth is 
curvature driven and an asymptotic regime during which diffusion
limited behavior is seen.

  The PFM, introduced and physically motivated 
by Langer \cite{Langer},
provides a theoretical
framework for many natural processes. It is designed
to treat situations where the relaxation dynamics of the 
order parameter
associated with the presence of a liquid or a solid is coupled to
the diffusion of heat released during the change of state.

An example is the growth of a solid nucleus
from its undercooled melt, a phenomenon encountered 
in rapidly solidifying materials, 
such as metals, where the growth is limited by the rate of transport of the 
heat of fusion away from the solid-liquid boundary \cite{Pelce}.
 As the heat released by the solid accumulates at the interface, it slows 
down the growth, because diffusion must act over a thicker and thicker 
region. 
This mechanism has also implications in the morphology of
the growing phases and is responsible for the instability of
a planar solid-liquid
interface with
respect to a perturbation of its shape; one realizes, immediately, that
a protrusion of the solid phase into the liquid advances 
faster than its neighboring regions, because 
it explores a region where the undercooling is greater, so its growth 
becomes faster. The solid-liquid surface tension eventually provides
the necessary balance and prevents the interface to be eroded by
fluctuations of very short wavelength \cite{Mullins,Muller}.

 Another very closely related problem, is the the growth of the solid
phase in multicomponent solutions, where one of the components 
is to be diffused away from the interface in order to form a stable
crystal \cite{Pelce}. In the present paper, we shall confine 
the discussion to the thermal
case for the sake of clarity, and
investigate the kinetic ordering of a spherical
version of the Phase field model. This study, extends
our previous investigations \cite{Noi,UMBM,Conti} 
to include a non vanishing order parameter. In our opinion it can be 
useful, because it
provides one the few models,
whose static properties can be obtained exactly in arbitrary dimensionality
and whose relaxation behavior can be analysed in great detail by means of
analytical and numerical methods.
We note that
in the field of ordering kinetics there exist 
only a few models for which the relaxation can be studied
without performing heavy
numerical calculations. In particular the late stage behavior of
processes with conserved dynamics
is hardly
observable numerically, due to computer limitations \cite{Conti,Collins}. 
 Besides the examples cited above, this study
may be of some help to treat analytically some models introduced
recently with the aim of describing irreversible aggregation phenomena
\cite{Maritan}.

 In the present paper we generalize the model, introduced 
previosly by two of us 
\cite{Noi,UMBM}, to the case of off-critical quenches, e.g. to
initial conditions corresponding to non-vanishing values of the fields.
The structure of the paper is the following. 
In order to make the paper self-contained
we have included two sections (I and II), where we recall
some basic notions, which lead to the thermodynamic derivation of 
the Phase field model \cite{Langer,Penrose,Wheeler,Kupferman}. 
and the construction of
the Lyapunov functional, from which the coupled equations
of motion of the PFM model can be derived . 
In section III, we state the Spherical Phase
field model and write explicitly 
the closed set of equations, which we
discuss qualitatively in section IV .
In section V comparisons with numerical
simulations in $d=2$ at zero temperature
are  illustrated. The predictions 
of the spherical model are confronted
with those of a more realistic scalar order parameter
Phase field model and the similarities and differences are stressed
in the conclusions.

\section{Thermodynamic preliminaries}
\label{sec:ThermP}
 Let us consider a material which
under suitable conditions of pressure and 
temperature can exist in two distinct thermodynamic phases, a liquid
and a solid. If the pressure is held constant
at the value corresponding to solid-liquid equilibrium 
and the temperature, $T$, is varied one can favor the solid phase
for $T<T_{M}$, or the liquid phase
for $T>T_M$, where $T_M$ is the melting temperature at which
the equilibrium first order transition takes place.
One usually calls  undercooled melt  a material brought 
below its melting temperature, but still in the liquid phase. 

We shall consider the situation $T<T_M$, which is
experimentally 
and technologically more interesting.
Below $T_M$ the value of the thermodynamic Gibbs potential 
of the solid phase is lower than that of 
the liquid phase, which is only metastable. 
A convenient way of studying the solidification process is to adopt
a phenomenological 
Ginzburg-Landau description by introducing a suitable crystalline
order parameter $\phi$ which assumes the conventional value 
$\phi_l$ in the liquid phase 
and $\phi_s$ in the solid phase.
One then employs a field theoretic free energy functional of the 
form:
\begin{equation}
\label{eq:p1}
F[\phi]=\epsilon\int d^d x\, \left[
        \frac{\xi^2}{2}(\nabla \phi)^2+f(\phi)
        \right]
\end{equation}
where $f(\phi)$ is a function of the order parameter $\phi$ 
with the property of having two minima of equal depth
at $\phi=\phi_l$ and $\phi=\phi_s$.

The constant $\epsilon$ has dimensions of energy/volume and is for
the moment arbitrary.
The gradient term represents the energy cost necessary to create an 
inhomogeneity in the system, the quantity $\xi$
has the dimension of length and is associated with the scale
over which a inhomogeneity in the system vanishes.
  Upon minimizing $F[\phi]$ with respect to 
$\phi$ and selecting the non uniform localized solution
of the variational Euler-Lagrange equation 
corresponding to the lowest value of the Gibbs
free energy
one obtains the surface tension $\sigma$  of the model 
which is proportional to the correlation length 
\begin{equation}
\label{eq:p4}
\sigma \sim \epsilon \xi
\end{equation}
and thus to the interface thickness. The numerical coefficient is of order 
$1$ and will be ignored, because it does not influence our discussion.

In order to include undercooling or overheating effects,
i.e., a temperature different from $T_M$, 
we introduce a dimensionless field 
\begin{equation}
\label{eq:u}
   u(\bbox{x})=\frac{c_p}{L}\,[T(\bbox{x})-T_M]
\end{equation}
proportional to the local temperature shift $(T(x)-T_M)$.
The constants $c_p$ and $L$ are respectively the 
specific heat at constant pressure and the latent heat of fusion per 
unit volume.
The local field $u$ acts as an external field, conjugate to
the crystalline order parameter $\phi$, favoring the solid phase
for $u<0$ and the liquid phase for $u>0$. For $u=0$ the two
phases coexist.

As usually done for first order phase transition, 
metastability is taken into account by eliminating $\phi$ in 
favour of $u$ via a Legendre transform. One then introduces the 
Gibbs potential 
\begin{equation}
\label{eq:g1}
 G[u] = F[\phi]-\lambda \epsilon \int d^d x\, u(\bbox{x})\, \phi(\bbox{x})
\end{equation}
where $\phi=\phi[u]$ is obtained form
\begin{equation}
\label{eq:g1p}
 \frac{\delta F[\phi]}{\delta \phi(x)} = \lambda \epsilon u(x)
\end{equation}
and $\lambda$ is a nondimensional parameter. 
A convenient way to
relate $\lambda$ to the known thermodynamic parameters 
is to consider the entropy difference between the pure uniform 
solid ($\phi=\phi_s$) and liquid ($\phi=\phi_l$) phases at the melting 
temperature. This is related to the latent heat by the relation
\begin{equation}
\label{eq:latent}
  S_l-S_s = \frac{L V}{T_M}
\end{equation}
where $S_l$ and $S_s$ are the entropies of the liquid phase and of
the solid phase, respectively, and $V$ the volume of the system.
By using the thermodynamic relation
\begin{equation}
\label{eq:termrel}
 \frac{\partial G}{\partial T}=-S, 
\end{equation} 
equation (\ref{eq:u}) and the expression (\ref{eq:g1}) 
for uniform fields, we get
\begin{eqnarray}
\label{eq:g4}
 \frac{LV}{T_M} &=& -\left.\frac{\partial G}{\partial T}\right|_{\phi_l,T_M}
                  +\left.\frac{\partial G}{\partial T}\right|_{\phi_s,T_M}
 \nonumber\\
                &=& \lambda\epsilon V (\phi_l-\phi_s) \frac{c_p}{L}
\end{eqnarray}
from which we obtain:
\begin{equation}
\label{eq:g7}
\lambda=\frac{L^2}{ \epsilon c_p T_M \Delta\phi}
\end{equation}
where $\Delta\phi=\phi_l - \phi_s$.

Next consider a solid spherical drop of radius $R\gg\xi$ 
immersed in an undercooled melt ($u<0$).
The Gibbs potential $G$ with 
the drop is -- see eq. (\ref{eq:g1}) --:
\begin{equation}
\label{eq:g5}
 G = G_0 + \lambda \epsilon u \frac{4\pi}{3}R^3 \Delta\phi + 
            4 \pi R^2 \sigma
\end{equation}
where the first term is the Gibbs potential $G$ without the droplet,
the second term is gain in replacing the liquid with solid in
the droplet, finally the third term is the cost in creating a surface 
separating the liquid and solid phases.
In equilibrium no energy is needed to create the droplet so
$G$ is stationary with respect to variation of $R$. By imposing
$\delta G=0$ we readily obtain the critical nucleation radius  $R_N$
\begin{equation}
\label{eq:g6b}
   R_N= \frac{d_0}{|u|}
\end{equation}
where 
\begin{equation}
\label{eq:g6}
   d_0= \frac{2 \sigma}{\Delta\phi \lambda \epsilon}
\end{equation}
is a capillarity length, which using the expression of $\lambda$,
eq.(\ref{eq:g7}), can be written as:
\begin{equation}
\label{eq:capl}
  d_0 = \frac{2\sigma c_p T_M}{L^2}.
\end{equation}
Finally from eqs. (\ref{eq:g7}) and (\ref{eq:g6}) it follows that 
we can write the dimensionless parameter $\lambda$ as the ratio
two length: 
\begin{equation}
\label{eq:g9}
\lambda=\frac{\xi}{d_0}
\end{equation}
where we have defined $\epsilon\,\xi = (2/\Delta\phi)\,\sigma$
[cfr. eq (\ref{eq:p4})].
We see that $\lambda$ is
small provided the interfacial thickness is much shorter than
the capillary length.

\section{The Phase field model}
\label{sec:PhField}
In this section we shall introduce relaxational dynamics into the model. 
 A large body of work in the area of dynamic phase transitions has 
been focused on the time-dependent Ginzburg-Landau (TDGL) model, because of its
capability of describing a variety of problems. In equilibrium the 
field $\phi(\bbox{x})$ minimizes the Gibbs potential $G$. Thus we assume that 
the approach to equilibrium is described by following equation:
\begin{eqnarray}
   \frac{\partial \phi(\bbox{ x},t)}{\partial t} &=& 
               - \Gamma_{\phi}\frac{\delta}{\delta\phi(\bbox{ x},t)}\, 
G[\phi,u] 
 \nonumber\\
              &=& - \Gamma_{\phi} [-\xi^2\nabla^2 \phi+ f'(\phi) -\lambda u]
\label{eq:cap2}
\end{eqnarray}
where the last equality is obtained using eq. (\ref{eq:g1}). If the 
field $u$ varies on time-scales much longer than those of $\phi$
it can be considered ``quenched'' and eq. (\ref{eq:cap2}) would be
the standard non conserved  TDGL equation, or model A.

In the Phase Field model, and in the absence of external sources,  
$u(\bbox{x},t)$ is assumed to evolve on time-scales 
of the same order of magnitude as those of $\phi$ towards an 
homogeneous configurations.
The time-evolution of $u$ is now coupled to that of $\phi$ and cannot
be neglected anymore. In fact,
when a piece of material solidifies it expels some heat and
the surrounding liquid melt warms up,  causing the average temperature
to increase. In turn  when a region of solid melts it adsorbs some heat
is adsorbed and the liquid becomes colder. 

As a consequence eq. (\ref{eq:cap2}) has to be supplemented with an 
equation for $u$.
The thermal field $u(\bbox{x},t)$ is subject to the Fourier equation 
of diffusion of heat plus an additional source term which 
represents the latent heat of solidification,
accompanying the appearance of the solid phase.

The energy balance requires that the latent heat released at the transition 
equates the temperature change of the melt multiplied the specific heat, i.e:
\begin{equation}
\label{eq:model0}
\frac{\partial u(\bbox {x},t)}{\partial t}
=D\nabla^2 u(\bbox {x},t)
-\frac{1}{\Delta \phi}\frac{\partial\phi(\bbox{x},t)}{\partial t}
\end{equation}
where $D$ is the thermal diffusivity and the last term in the
right-hand side is the amount of material which crystallises
per unit time and thus proportional to the heat released 
during the first order transition \cite{Langer}.
The coefficient $\Delta \phi$ guarantees
the correct energy balance. Notice that the last term represents a source
of heat when ${\partial\phi(\bbox{x},t)}/{\partial t}$ is negative, i.e.
when the system solidifies, or a sink when it melts, positive
${\partial\phi(\bbox{x},t)}/{\partial t}$.
In other words, since we are considering a
closed system the total amount
of solid produced is proportional to the change of the average
temperature of the system.

The two dynamical equations (\ref{eq:cap2}) and (\ref{eq:model0})
can be obtained from a unique Lyapounov functional ${\cal F}$, which
plays the role of the time dependent Ginzburg-Landau potential in the 
present problem. In order to establish the form of ${\cal F}$ 
we perform the transformation 
\begin{equation}
\label{eq:bigU}
  U = u + \frac{\phi}{\Delta \phi}
\end{equation} and eliminates $u$ in favor of the new field 
$U$. One can then write eqs. (\ref{eq:cap2}) and (\ref{eq:model0}) as
\begin{eqnarray}
      \frac{\partial \phi(\bbox{x},t)}{\partial t} &=&
     -\Gamma_{\phi}\left.\frac{\delta {\cal F}}{\delta \phi({\bbox x},t)}
       \right|_{U}
\label{eq:quattro}
  \\
\frac{\partial U(\bbox{x},t)}{\partial t} &=& D\,
               \nabla^2\left.\frac{\delta {\cal F}}{\delta U(\bbox{x},t)}
                       \right|_{\phi}
\label{eq:cinque}
\end{eqnarray}
with the Lyapounov functional \cite{Collins,Conti}:
\begin{equation}
\label{eq:tre}
{\cal F}[\phi,U]=\int d^dx\, \left[\frac{\xi^2}{2}(\nabla \phi)^2+
           f(\phi)+\frac{\lambda \Delta \phi}{2} 
                \left(U-\frac{\phi}{\Delta \phi}\right)^2
                           \right].
\end{equation}
Note that the dynamics of $U$ is conserved.
When the temperature field vanishes, i.e.,  $U=\phi/\Delta \phi$,
the functional ${\cal F}$ has
two equivalent minima, corresponding to two spatially uniform
solutions: the uniform solid and liquid
phases. 
In general eqs. (\ref{eq:quattro})-(\ref{eq:tre}) generate
a complex dynamical behavior which
has been the object of some studies. 

In the long time limit we may expect that while $u$ becomes 
homogeneous, the crystalline field $\phi$ roughly assumes only 
the two values $\phi_l$ and $\phi_s$. If this is the case, from the
knowledge of $U$ and $u$ we can compute the fraction of volume 
occupied by the two phases. Indeed we can write 
$\overline{\phi} = x_s\phi_s + x_l\phi_l$, where $x_s$ and $x_l$ are the fraction
of volume occupied by the solid and liquid phase, respectively, and
the overbar denotes spatial average.
 From eq. (\ref{eq:bigU}) and the condition $x_s+x_l=1$, we get
\begin{equation}
	\begin{array}{ll}
   x_l &= -\frac{\displaystyle \phi_s}{\displaystyle\Delta\phi} +
              \overline{U} - \overline{u}
   \\
   x_s &= \frac{\displaystyle \phi_l}{\displaystyle\Delta\phi} - 
              \overline{U} + \overline{u}
	\end{array}
\label{eq:xli}
\end{equation}
where $\overline{u}\simeq u$ is the asymptotic value, while $\overline{U}$ 
is the initial value being its dynamic conserved. From this it follows that if
the asymptotic value of $u$ is zero, i.e., the system relaxes towards
a two phase coexistence, the fraction of volume occupied by each phase
is determined only by the initial value of $\overline{U}$.
 We also notice that when the system starts with an undercooling 
$\overline{u}=-1$, and order parameter $m(t=0)=1$, 
i.e. $\overline{U}$, the latent heat produced is just enough to heat the 
melt at the final equilibrium temperature, $\overline{u}=0$.
In such a case the final
volume fraction of the solid is simply $x_s=1/2-U$, and attains its
maximum for $U=-1/2$.

The functional 
${\cal F}$ decreases with time
as it can be shown using the equations (\ref{eq:quattro}) and (\ref{eq:cinque}):
\begin{eqnarray}
\label{eq:glo1}
 \frac{d {\cal F}}{dt} &=&
    \int d^dx\,\left[
               \frac{\delta {\cal F}}{\delta \phi({\bbox x},t)}
                \frac{\partial \phi(\bbox{x},t)}{\partial t}
             + \frac{\delta {\cal F}}{\delta U({\bbox x},t)}
                \frac{\partial U(\bbox{x},t)}{\partial t}
                \right]
 \nonumber\\
                       &=& -\int d^dx\, \left[
                 \Gamma_{\phi}\,
                   \left(\frac{\delta {\cal F}}{\delta \phi({\bbox x},t)}\right)^2
                +D\,
                   \left(\frac{\nabla \delta {\cal F}}{\delta U({\bbox x},t)}\right)^2
                                       \right] \leq 0.
\end{eqnarray}

So far we have discussed purely deterministic evolution
of the order parameter and of the thermal field. Noises can be added
to both equations to represent the effect of short wavelength 
fluctuations; in this case eq. (\ref{eq:glo1}) does not hold.

\section{Spherical Phase Field Model}
\label{sec:spherconst}
The choice of the local function $f(\phi)$ is somehow arbitrary, as long as 
the general property 
\begin{equation}
\label{eq:genprop}
  \lim_{\phi\to\pm\infty} f(\phi) = +\infty, \qquad
   \mbox{\rm and two equal minima for}\ \phi=\phi_s\ 
\mbox{\rm and}\ \phi=\phi_l
\end{equation}
are satisfied.
A largely used form for $f(\phi)$ is
\begin{equation}
\label{eq:locf}
  f(\phi) = -g\,\left(\frac{\phi^2}{2} - \frac{\phi^4}{4}\right)
\end{equation}
which is even and has two equal minima at $\phi=\phi_s=-1$ and $\phi=\phi_l=1$, 
so that $\Delta\phi=2$.
The parameter $g$ gives the strength of the local constraint. In the limit
of large positive value of $g$ the field $\phi$ can take only the
values $\phi_s$ and $\phi_l$ (Ising-like variables).

The phase field model described by equations (\ref{eq:quattro})-(\ref{eq:tre})
contains all the relevant ingredients
necessary to describe the phase separation occurring in solid forming melts. 
However, due to the local nonlinear terms contained in the function $f(\phi)$
the solution of the dynamical equations (\ref{eq:quattro})-(\ref{eq:tre})
are far too difficult and are known only for some special situations. 
In the general case the known results follow from dimensional arguments.

To overcome this difficulty, an alternative strategy is to 
modify the model into a simpler one, yet maintaining the general properties.
This can be achieved by replacing the local quartic term in (\ref{eq:locf}):
\begin{equation}
 \label{eq:glof}
        \int d^dx\, \phi(\bbox{x})^4 \to 
      \frac{1}{V} \left[\int d^dx\, \phi(\bbox{x})^2\right]^2.
\end{equation}
This kind of constraint is much softer 
than (\ref{eq:locf}) since it does not act on each site, but globally over the
whole volume. In the following we shall denote it as {\it globally constrained 
model} or {\it spherical model} \cite{Berlin,Zann,Ciuchi} in 
contrast with the model 
(\ref{eq:quattro})-(\ref{eq:tre}) where the constraint is local
\cite{sferico,Stanley}.

The price one pays for this change is the
loss of sharp interfaces between two coexisting phases. 
As Abraham and Robert\cite{Abraham} showed longtime ago the spherical model
in zero external field displays two ordered phases below the critical
temperature, but no phase separation. Equivalently 
one can say that a planar interface between two
coexisting phases is unstable, due to the presence of long-wavelength
excitations analogous to spin-waves, an instability much stronger
than the one due to the presence of capillary waves in the
scalar order parameter case \cite{Weeks,Gyorffy}.
As a consequence, while this choice  is
very convenient for analytic calculations, it changes
the structures  of the non uniform
solutions in the static limit.
Nevertheless, 
in spite of this fact, the model has a rich phenomenology as we
shall see below and the approach to equilibrium remains highly non trivial. 

  From equations (\ref{eq:tre}), (\ref{eq:locf}) and (\ref{eq:glof}), the
potential ${\cal F}$ for the spherical model reads:
\begin{equation}
\label{eq:hh}
     {\cal F}[\phi, U]= \int d^dx\, \left[
                \frac{\xi^2}{2}\,(\nabla \phi)^2
              + \frac{1}{2}\,\left(\frac{\lambda}{2} - g\right)\,\phi^2
              + \lambda U^2
              - \lambda U\phi
                                    \right]
              + \frac{g}{4V}\left( \int d^dx\, \phi^2 \right)^2,
\end{equation}
which substituted into:
\begin{equation}
\label{eq:Lang1}
   \frac{\partial \phi(\bbox{ x},t)}{\partial t} = 
  - \Gamma_{\phi}\frac{\delta}{\delta \phi(\bbox{ x},t)}\, {\cal F}[\phi, U] 
               + \eta(\bbox{x},t)
\end{equation}
\begin{equation}
\label{eq:Lang2}
   \frac{\partial U(\bbox{ x},t)}{\partial t} = 
          D\nabla^2 \frac{\delta}{\delta U(\bbox{ x},t)}\, {\cal F}[\phi,U] 
               + \xi(\bbox{x},t)
\end{equation}
determines the time evolution of the fields $\phi$ and $U$.
We added to the evolution equations a noise term to simulate the
effect of short wavelength fluctuations.
The two fields $\eta$ and $\xi$ are independent Gaussian fields
with zero mean and two-point correlations:
\begin{eqnarray}
           \langle\eta(\bbox{x},t)\, \eta(\bbox{x'},t')\rangle &=& 
           2\,T_f\,\Gamma_{\phi}\, \delta(\bbox{x-x'})\, \delta(t-t')
\label{eq:noise1}
   \\
           \langle\xi(\bbox{x},t)\, \xi(\bbox{x'},t')\rangle &=&
            -2\,T_f\,D \,\nabla^2 \delta(\bbox{x-x'})\, \delta(t-t')
\label{eq:noise2}
   \\
           \langle\eta(\bbox{x},t)\,\xi(\bbox{x'},t')\rangle &=& 0
\label{eq:noise3}
\end{eqnarray}
where $T_f$ is the temperature of the 
final equilibrium state whereas $D$ and $\Gamma_{\phi}$ are the
kinetic coefficients appearing into eqs. (\ref{eq:Lang1}) and (\ref{eq:Lang2}).

It is useful to separate out the spatially uniform component 
of fields $\phi$ and $U$. Thus introducing the Fourier component
of the fields we have:

\begin{equation}
 \begin{array}{ll}
 \label{eq:FouriSa}\displaystyle
 \phi(\bbox{x}) = \sum_{\bbox{k}}\, \phi(\bbox{k})\, e^{i\bbox{k}\cdot\bbox{x}}
\\
 \label{eq:FouriSb}\displaystyle
 \phi(\bbox{k}) = (1/V)\,\int d^dx\, \phi(\bbox{x})\, e^{-i\bbox{k}\cdot\bbox{x}}
 \end{array}
\end{equation}
and

\begin{eqnarray}
    \phi(\bbox{k},t) &=& m(t) \delta_{k,0}+\delta \phi(\bbox{k},t) 
 \label{eq:phi1}
     \\
    U(\bbox{k},t) &=& Q(t) \delta_{k,0}+\delta U(\bbox{k},t) 
\label{eq:U1}
\end{eqnarray}
where both $\delta\phi$ and $\delta U$ are zero for $k=|\bbox{k}|=0$, 
$m(t) = \overline{\phi}(t)$, $Q(t) = \overline{U}(t)$.

To study the behaviour at finite temperature $T_f$ it is also useful 
to introduce the equations of motion for the three
equal-time real space connected correlation functions
 $C_{\phi\phi}(r,t)=\langle\phi(R+r,t) \phi(R,t)\rangle_c$, 
 $C_{\phi U}(r,t)=\langle\phi(R+r,t) U(R,t)\rangle_c$ and 
 $C_{UU}(r,t)=\langle U(R+r,t) U(R,t)\rangle_c$, 
whose Fourier transforms are the structure functions.
The average $<\*>$ is over the external noises $\eta$ and $\xi$ and initial 
conditions. 

Due to the special form of the non-linear term in the equation
of motion the set of evolution equations for
the averages $m(t)$, $U(t)$ and
the correlation functions is closed. Indeed, in the Fourier space
these read:
\begin{eqnarray}
 \frac{\partial\phi(\bbox{k},t)}{\partial t} &=& 
       F_{\phi}(\bbox{k}) + \eta(\bbox{k},t)
 \label{eq:ophi1}
     \\
 \frac{\partial U(\bbox{k},t)}{\partial t} &=& 
       F_{U}(\bbox{k}) + \xi(\bbox{k},t)
\label{eq:oU1}
\end{eqnarray}
where $F_{\phi,U}$ are the Fourier transforms of the 
first term on the right-hand sides of eqs. (\ref{eq:Lang1}) and
(\ref{eq:Lang2}). From (\ref{eq:hh}) we have
\begin{eqnarray}
   F_{\phi}(\bbox{k}) &=& 
                   M_{\phi\phi}(k,t)\, \phi(\bbox{k},t)
                 + M_{\phi U}(k,t)\,    U(\bbox{k},t)
\label{eq:phi}
   \\
   F_{U}(\bbox{k}) &=& 
                   M_{U \phi}(k,t)\, \phi(\bbox{k},t)
                 + M_{U U}(k,t)\,    U(\bbox{k},t).
\label{eq:U}
\end{eqnarray}
where the matrix elements are given by
\begin{equation}
\label{eq:matrix}
\begin{array}{ll}
          M_{\phi\phi}(k,t) =& -\Gamma_{\phi}[\xi^2 k^2+r+ gm^2(t)+ gS(t)], \\
          M_{\phi U}(k,t)=& \Gamma_{\phi} \lambda \\
          M_{U \phi}(k,t)=& D \lambda k^2, \\
          M_{U U}(k,t) =&- 2 D \lambda k^2.
\end{array}
\end{equation}
where $r=(-g+\lambda/2)$
and, in the limit $V\to\infty$ the quantity $S(t)$ is the integrated 
${\bf \phi}$-structure function
\begin{eqnarray}
  S(t) &=& \frac{1}{V} \int\, d^dx \langle\phi(\bbox{x},t)\,
                                              \phi(\bbox{x},t)\rangle
 -m^2(t)
 \nonumber\\ 
       &=& \sum_k \langle\phi(\bbox{k},t)\,
                                              \phi(-\bbox{k},t)\rangle_c
\label{eq:St}
\end{eqnarray}
and $m(t) = \langle\phi\rangle$.

As a consequence we have
\begin{eqnarray}
\label{eq:op}
\frac{\partial m(t)}{\partial t} &=& M_{\phi\phi}(0,t) m(t) +M_{\phi U}(0,t)  Q(t) 
\\
\label{eq:up}
\frac{\partial Q(t)}{\partial t}&=&0
\end{eqnarray}
expressing the conserved nature of the field $U$, and
\begin{eqnarray}
\label{eq:c11}
\frac{1}{2}\,\frac{\partial}{\partial t}\, C_{\phi\phi}(k,t) &=& 
M_{\phi\phi}(k,t) C_{\phi\phi}(k,t) \nonumber \\
&+&M_{\phi U}(k,t) C_{\phi U}(k,t) +\Gamma_{\phi}T_f
\end{eqnarray}
\begin{eqnarray}
\label{eq:c12}
\frac{\partial}{\partial t}\, C_{\phi U}(k,t)&=& 
M_{U\phi}(k,t) C_{\phi \phi}(k,t) \nonumber \\
&+&\bigl[M_{U U}(k,t)
+M_{\phi\phi}(k,t)\bigr]C_{\phi U}(k,t) \nonumber \\ 
&+&M_{\phi U}(k,t) C_{U U}(k,t) 
\end{eqnarray}
\begin{eqnarray}
\label{eq:c22}
\frac{1}{2}\,\frac{\partial}{\partial t}\, C_{U U}(k,t)&=& 
M_{U\phi}(k,t) C_{\phi U}(k,t) \nonumber \\
&+&M_{U U}(k,t) C_{U U}(k,t) + D T_f k^2.
\end{eqnarray}

We note that a closure at the same level
would have been obtained in the framework of a Hartree approximation
for the model with local constraint described by 
eqs. (\ref{eq:quattro})-(\ref{eq:cinque}).
However, within the present model the eqs. (\ref{eq:op})-(\ref{eq:c22}) 
are exact and not the result of a approximate decoupling of the correlations.

\section{Long time behaviour}
\label{sec:longtb}
In this section we shall discuss the behaviour of the spherical Phase Field 
model for long times, i.e., $t\to\infty$. The results will be compared with 
those of direct numerical simulation in the next section.

We assume that at the initial time we have an undercooled liquid with some 
supercritical solid seeds. This means that $\overline{u} < 0$ while
$m=\overline{\phi}$ lies in the interval $(0,1)$.
For a generic initial configuration, the undercooled liquid is not in 
equilibrium. Thus  at the initial stage the relaxation of the field $\phi$ 
is only slightly modified by the dynamics of the slower field $U$, which 
can be considered ``almost quenched''. During this stage the size of the
solid seeds grow with time, while the maximum of the structure function
is located at $k=0$ and grows with time. If this regime is long enough,
one can recognize a typical non conserved order parameter dynamics (NCOP)
domain growth proportional to $t^{1/2}$.

This kind of behaviour persists until the typical size of the domains
reaches that associated to the conserved $U$ field. At this time the
dynamics of the two fields becomes strongly correlated, and the 
conserved order parameter dynamics eventually dominates. As a consequence the
$\phi$ field slows down, since the coupling with the conserved $U$ field
acts as an additional constraint, while $m(t)$ becomes nearly constant, and
equal to the asymptotic value, i.e.,  $m\sim\overline{\phi}_{\infty}$.

The cross-over time $t_c$ can be readily estimated from dimensional arguments.
An inspection of equations (\ref{eq:ophi1})-(\ref{eq:oU1}) reveals that 
under a suitable transformation of parameters, in which $\lambda t\to t$, 
the dimensionless parameter $\lambda$ can be traced out. 

This means that the crossover time, $t_c$, to the conserved
dynamics is also of order  $ \sim 1/\lambda$. 
It can be shown that the Langevin equations (\ref{eq:ophi1})-(\ref{eq:oU1}) 
obey detailed balance \cite{UMBM}, and that the stationary probability density
is
\begin{equation}
\label{eq:StaPD}
 P_{st}[\phi,U] \propto \exp\left(-\frac{1}{T_f}\,{\cal F}[\phi,U]\right).
\end{equation}
Using this we easily get the equilibrium $\phi$$\phi$ correlation
function, which  reads
\begin{equation}
\label{eq:Cppe}
 \langle\phi\,\phi\rangle = \frac{T_f}{\xi^2 k^2+r+gS+gm^2-\lambda / 2}.
\end{equation}
The appearance of an ordered phase, with $m\not=0$ for temperatures $T_f$,
below the critical temperature $T_c$, is revealed by the divergence of the
$k=0$ mode. This implies that the equilibrium (
i.e. when $t\to\infty$) value of $m$
must satisfy the following equation:
\begin{equation}
\label{eq:condeq}
  r + gS + gm^2 - \lambda/2 = 0.
\end{equation}
On the other hand, from the equation of motion (\ref{eq:op}) 
$m$ must be solution of:
\begin{equation}
\label{eq:conddy}
 rm + gm^3 + gSm - \lambda Q = 0.
\end{equation}
The simultaneous solution of eqs. (\ref{eq:condeq})-(\ref{eq:conddy})
requires:
\begin{equation}
\label{eq:mas}
 m = 2Q
\end{equation}
which via eq. (\ref{eq:bigU}) together with $\Delta \phi=2$
implies $\overline{u}=0$.
This means that the
system relaxes towards a nontrivial phase coexistence state,
with non vanishing order parameter and diverging small $k$ fluctuations.

One sees that
the condition (\ref{eq:mas}) can be satisfied only for $-1/2\le Q \le 1/2$. 
If $Q$ lies outside this interval the system does not relax to a
mixed phase, but instead settles in a spatially uniform state
without zero modes.
Indeed, in this case $S(t)$ vanishes and $m$ relaxes for long times to 
the value given by, cfr. (\ref{eq:op}),
\begin{eqnarray}
\label{eq:conddy2}
      rm &+& gm^3 - \lambda Q = 
\nonumber\\
      -gm&(& 1-m^2) - \lambda\overline{u} = 0.
\end {eqnarray}
We note that eq. (\ref{eq:conddy2}) is equivalent to say that a
a spatially uniform field $\phi(x)$ is a stable minimum of the
the potential ${\cal F}$. For values large enough of
$|\overline{u}|$ equation (\ref{eq:conddy2})
has a single solution, positive for  $\overline{u}>0$ and negative 
for $\overline{u}<0$. These correspond to a liquid phase above the melting
temperature, i.e., positive $m$ and $\overline{u}$, and a solid phase below
the melting temperature, negative $m$ and $\overline{u}$.
As $|\overline{u}|$ decreases two additional solutions eventually appear. 
One of these is
unstable while the other represents a metastable spatially uniform state:
i.e. solid above the melting 
temperature ($\overline{u}>0$) and liquid below the melting temperature
($\overline{u}<0$) . 
These solutions, however, are unstable 
against fluctuations, indeed the presence of $S(t)$,
which vanishes only for $t\to\infty$ prevents the dynamics to reach these 
metastable states. Therefore for any 
value of $\overline{u}$ the physical solution
of eq. (\ref{eq:conddy2}) is the most negative one for $\overline{u}<0$ and
the most positive one for $\overline{u}>0$.
For small $\overline{u}$ we have
$m= (1 - 2(\lambda/g)\overline{u})\,{\rm sign}\overline{u}$.

A more detailed analysis of the  approach to equilibrium can be done employing
a quasilinearization procedure, i.e. we assume that
the quantity 
$ R(t) = r + gm(t)^2+gS(t)$ can be treated as a constant, along different
pieces of the trajectory \cite{Noi}.
 One can verify at posteriori that the assumption
is valid and leads to useful predictions. Since the behaviour at $T_f=0$ is 
representative of the entire dynamics in the ordered phase when $T_f<T_c$, 
we also set $T_f=0$, without loosing relevant information.

Assuming the quantity $R(t)$ to be nearly constant eqs. (\ref{eq:phi1}) and 
(\ref{eq:oU1}) become a linear system whose solution has the form:
\begin{equation}
\label{eq:phib}
\begin{array}{ll}
  \phi(k,t) &=   c_{\phi}^{+}(k)\, e^{\omega_{+}(k)t}
               + c_{\phi}^{-}(k)\, e^{\omega_{-}(k)t} \\
  U(k,t)   &=   c_{U}^{+}(k)\, e^{\omega_{+}(k)t} 
              + c_{U}^{-}(k)\, e^{\omega_{-}(k)t}
\end{array}
\end{equation}
where $\omega_{+}(k)$ and $\omega_{-}(k)$ are the eigenvalues of
the $M$ matrix,
\begin{equation}
\label{eq:eigen}
   \omega_{\pm}(k)=\frac{1}{2}\,\left[-\Gamma_{\phi}
(\xi^2 k^2+R)-2 D    \lambda k^2
                   \pm \sqrt{[\Gamma_{\phi}(\xi^2 k^2+R)+ 2 D 
                    \lambda k^2]^2+
                             4\,\Gamma_{\phi} D \lambda^2 k^2}
                                \right]
\end{equation}

For time $t\gg 1$ the dynamical behavior of the solution is
determined by the larger eigenvalue $\omega_{+}(k)$
For large values of $k^2$ the eigenvalue $\omega_{+}(k)$ decreases as
$-k^2$, thus to discuss the behavior of the solutions after the initial 
transient
a small $k$ expansion of $\omega_{+}(k)$ is sufficient.
The form of this expansion depends on the sign of $R(t)$. When $R(t)$ is
negative 
the appropriate expansion of $\omega_{+}(k)$ is:
\begin{equation}
\label{eq:omp}
 \omega_{+}(k)=\Gamma_{\phi}|R|-
                \left[\Gamma_{\phi}\xi^2-D\frac{\lambda^2}{|R|}\right]k^2
\end{equation}
Notice that in thie regime where eq. (\ref{eq:omp}) 
is valid there is a competition
between the curvature term $\Gamma_{\phi}\xi^2$, 
which represents the driving force of the dynamics of
the pure  Model A, and the term $D\lambda^2/|R|$ due to the coupling to the
heat diffusion.
 For $R(t)>0$, the representation eq. (\ref{eq:omp}) breaks down
and one one must instead consider the expansion:
\begin{equation}
\label{eq:dispersion2}
 \omega_{+}(k)=2 D \lambda \left( \frac{\lambda}{2 R}-
 1 \right)k^2 - c_4 k^4
\end{equation}
where $c_4$ is a positive coefficient.

Since the value of $R(t)$ as well as its sign change along the trajectory, 
one must employ either eq. (\ref{eq:omp})  or eq. (\ref{eq:dispersion2}) 
depending on the stage of the growth process. One the other hand, this kind of 
analysis is not applicable in the crossover
region where $m(t)$ and $S(t)$ vary too rapidly
and $R(t)\sim 0$, but this fact does
not invalidate our findings.

All the relevant behaviors can be classified according to the value of the 
conserved field $Q$; to this 
purpose one must treat separately the cases $Q>0$
and $Q<0$.

Let us consider first the case  $Q<0$
and assume $m(t=0) \sim 1$, i.e. the system initially
is formed by an undercooled 
liquid. 
In the initial stage the fields $\phi$ and  $U$ are nearly uniform
and characterized by small fluctuations. In the evolution equation
for $m$ , $U$ plays the role of a constant field of value $Q$, hence
\begin{equation}
\label{eq:mm}
 \frac{\partial m}{\partial t}=-\Gamma_{\phi}\,\bigl[R(t)m(t)-\lambda Q\bigr]
\end{equation}
describes the relaxation of $m$ in a static field $Q$. During 
such a stage the system 
relaxes towards the nearest fixed point 
which makes the right-hand side of 
(\ref{eq:mm}) to vanish.
Equating to zero $\partial m / \partial t$ we get the relation
\begin{equation}
\label{eq:static}
 R(t) = \lambda \frac{Q}{m}
\end{equation} 
which is negative for $m>0$ and $Q<0$, so that
the relevant expansion is (\ref{eq:omp}). Thus the liquid phase 
is unstable as it can be seen from
eq. (\ref{eq:omp}); in fact, as long as $m(t)$ remains positive the
system develops strong fluctuations about the uniform mode, $k=0$.
Such a regime lasts for a time of order $t_c\sim 1/\lambda$, after which
the fluctuations eventually drive the system towards negative 
values of the order parameter
as to reduce the free energy cost. 

In the successive stage the system becomes 
prevalently solid, and this is signalled
by the change of sign of $m$.
Also $R$ changes sign so the expansions (\ref{eq:omp}) 
becomes invalid. 
 However, for times $t\gg 1/\lambda$, long after the transition, the
evolution of $m$ slows down again, i.e.
$\partial m / \partial t \simeq 0$ and $R>0$, 
so that (\ref{eq:dispersion2}) is appropriate.
 Using the result eq.(\ref{eq:static}) 
the relevant form of $\omega_{+}(k)$ is:
\begin{equation}
\label{eq:dispersion2n}
 \omega_{+}(k)=D \lambda \left( \frac{m}{2 Q}-
                             1 \right)k^2 - c_4 k^4.
\end{equation}
Such expression shows that whenever $m/2 Q<1$
all modes of finite wavevector are  damped and one cannot observe
growing modes: the system relaxes towards a spatially uniform state.

On the other hand if $m/2 Q>1$ fluctuations
are large up to a finite wavevector, because the 
fastest growing mode is located at a finite $k$. This case corresponds to 
a phase spatially non uniform with large fluctuations. Asymptotically
$m=2Q$ and the peak position $k_m$, moves towards vanishing values of $k$.
This scenario is typical of the conserved order parameter (COP) dynamics.
Indeed, as done in Ref. \cite{Noi,UMBM}, it can be shown that in this regime
the dynamics exhibits multiscaling \cite{Noi}.

In Fig. \ref{fig:mdot} we report the behavior of $ \partial m / \partial t$ 
versus the order  parameter $m$.
For small negative values of  $\lambda Q$, 
$\partial m / \partial t$ nearly touches
the horizontal axis, and the evolution of $m$ becomes very slow.

Let us now turn to the $Q>0$ 
case. In this case from (\ref{eq:static}) follows that
$R(t)>0$, so that the relevant expansion is (\ref{eq:dispersion2n}).
In analogy with the $Q<0$ case,
from (\ref{eq:dispersion2n}) we conclude that 
if $Q>1/2$, the phase $m\sim 1$ is
stable and the final state is pure, i.e.
is characterized by small spatial fluctuations.

For smaller value of $Q$, i.e.,  $0<Q<1/2$
the coefficient of the term of order $k^2$ of the eigenvalue $\omega_{+}$
is positive and the peak of $C_{\phi\phi}(k,t)$ is located at
a finite wavevector. This causes an instability of the initial pure
phase $m\sim 1$ and the appearance of a mixed phase, characterized by
a lower, but still positive, value of $m(t)$.

\section{Numerical results} 
\label{eq:NumRes}
In what follows we shall consider the zero temperature
dynamics, because it is known to be representative of the subcritical
behavior. We have compared the numerical results with the predictions
of the previous section and found good agreement.
 
We have solved numerically the equations of motion 
(\ref{eq:ophi1})-(\ref{eq:oU1}) in two dimensions by using a simple Euler 
second order algorithm and a discretization of the integrals
on a $N\times N$ bidimensional lattice. We used periodic boundary conditions.
The parameters employed are
\begin{equation}
\label{wq:paraNum}
   \begin{array}{lcl}
         \xi = 1, &\qquad& g = 1 / 2\epsilon^2 \\
         \lambda = 4\Lambda / \epsilon^2 &\qquad& \Gamma_{\phi} = 1 /\alpha \\
         D = \epsilon^2 / 8\Lambda &\qquad&\\
   \end{array}
\end{equation}
with $\epsilon = 0.005$, $\Lambda = 0.1$ and $\alpha = 10$.
The time step used is  
$\Delta t = 2\times 10^{-5}$ and a lattice space $\delta x = 0.01$. 
Different values of $N$ were used, here we report the results for $N=256$.

The system was initially prepared in an non uniform 
initial state formed by $70\%$ of undercooled liquid and the remaining
$30\%$ of seeds of solid  randomly distributed. 
This ensures that at the initial time the  order parameter $m(t=0)$ is positive 
and the correlations are small. As long as these two conditions are met, 
the results are not too sensitive to the initial solid/liquid fraction.
The thermal field $u$ was taken uniform, 
$u_{ij}(t=0) = \overline{u}(t=0)$, with both positive and negative values.
According to eq. (\ref{eq:bigU}), the
conserved field $U$ was chosen to be equal
to
\begin{equation}
\label{eq:U0} 
    U_{ij}(t=0) =  \overline{u}(t=0) + \frac{1}{2}\,\phi_{ij}(t=0).
\end{equation}

The model exhibits two different long-time regimes as far as the 
temperature field is concerned. 
The snapshots of the field $\phi$ indeed reveal that changing from 
positive values of the conserved field $Q$ to negative values the 
long-time morphology changes. 
For $Q>0$ one observes solid drops immersed in a liquid and $u\le 0$,
whereas for $Q<0$ liquid drops become trapped in a solid matrix and $u\ge 0$.
This is in agreement with the general results of thermodynamics.
For quenches with positive $Q$ the minority phase is the solid, 
thus we expect the solid drops to have positive curvature. Indeed,
relating the curvature ${\cal{K}}$, to the 
temperature field we obtain from the Gibbs-Thompson
condition $u=-d_0 {\cal{K}}$ \cite{Pelce}
a negative value of $u$ 
for $Q>0$. On the contrary, upon crossing the $Q=0$ line, the
solid becomes the majority phase and $u$ change sign, since the
curvature relative to the solid is negative.

In Fig. \ref{fig:figG1} we show the behaviour of the order parameter
$m$ as function of time for the four distinct regimes. 
In all cases at the initial 
time we have $m=0.54$. We note that in the case $|Q|>1/2$ the system evolves towards 
a spatially uniform state, which is a liquid above the melting temperature
for $Q>1/2$ and positive initial $\overline{u}(t=0)$, case (a) in figure, or
a solid below the melting temperature for $Q<1/2$ and negative initial 
$\overline{u}(t=0)$, case (d) in figure. This state minimizes the potential 
${\cal F}$.  

In the case $|Q|<1/2$ the system evolves towards a phase 
equilibrium state at the
melting temperature: 
we have for $t\to\infty$ $u=\overline{u}=0$ and
$m=2Q$, in agreement with the analytical results of the previous section.

When the systems starts from an undercooled state, i.e. $u<0$ at the
initial time -- cases (b), (c) and (d) in Fig. \ref{fig:figG1} -- 
despite the fact the liquid state is unfavorable, $m$ increases  for short times.
The system then evolves towards a liquid state, $m$ tends to saturate 
to a fixed value $\sim 1$ and the fluctuations are small. This state is 
however unstable with respect to fluctuations, and indeed after a time $t_c$ 
we observe a transition towards the asymptotically stable state, which depends
on the value of $Q<1/2$.

The time $t_c$ is a decreasing function of $\lambda$, as can be seen
in Fig. \ref{fig:figA1}, where we report $m$ as a function of time
for $-1/2 < Q < 0$ and variuous values of $\lambda$. In Fig. \ref{fig:figA2}
we report $t_c$ versus $\lambda$ for the curves of Fig. \ref{fig:figA1}.
The line is the theoretical prediction $t_c\sim 1/\lambda$.
In this figure we defined $t_c$ as the time such that $\min_{t}m(t)=2Q$. Other
definitions are possible, for example when $m(t)=0$, or any other fixed
value. All these definitions leads to the same scaling with $\lambda$.

The scenario here depicted remains valid also for the local constraint case.
This has been checked by numerically integrating the appropriate equations.
The main difference between the 
local and the global case shows up at morphological 
level, as can be seen in the snapshots of Figs \ref{fig:figG2} and \ref{fig:figG3}. 
As one can see while the phase field model with local constraint has 
sharp domain wall the model with the global constraint presents smoother interfaces.
We stress that despite this difference the circularly averaged correlation 
functions and the structure factors for two cases are quite 
similar. In Figs. \ref{fig:figG4} (global) and \ref{fig:figG5} (local) 
we report the circularly averaged $\phi\phi$ correlation for the situations of figures 
\ref{fig:figG2}-\ref{fig:figG3} (circles), as well as for the case $|Q|>1/2$
(diamonds). The average radius of the drops is identified by the first zero of 
correlation functions.

While the simulations confirm the scenario described in Sec. \ref{sec:longtb}, 
we cannot extract the power law exponent predicted by the non conserved 
ordered parameter dynamics at short times, as well as those of the
conserved order parameter dynamics for long times. Indeed finite size effects
prevents us from reaching the conserved regime.

\section{Conclusions}
\label{sec:conc}
We have studied a model which reproduces many of the features 
which render appealing the scalar Phase field model and analysed 
its equilibrium and off-equilibrium properties.
We transformed the original scalar model into a model with
global couplings, more amenable to analytic investigations.

In such a description the temperature shift from the coexistence temperature
plays the role of an annealed field, which changes during the process
and settles to a value determined by energetic considerations. Its
dynamics is slow compared with that of the order parameter field
and the latter becomes eventually slaved by the first.

The long time state can be either a pure state with vanishing correlations,
in the limit $T_f=0$, or a mixed state with large spatial
fluctuations in the order parameter. The type of equilibrium reached depends 
on the initial value of the spatial average $Q$ of the field $U$.
Indeed, in the case $-1/2 < Q <1/ 2$,  the system
shows a tendency towards
separation into two phases in proportions  given by the rule 
$m=2 Q$ and one observes drops of the minority phase in a sea 
formed by the majority phase. At the same time the thermal field $u$ vanishes,
indicating that $T=T_M$ in the whole volume.
The number of the drops decreases
with time as to minimize the free energy of the system, but
for long times the total amount of solid remains fixed because
the heat released by a  growing solid
drop can only be adsorbed by a shrinking solid drop. 
As a result 
the solid order parameter $\phi$ 
becomes nearly conserved being mediated by the conserved heat field.
At this stage the dynamics of the crystalline order parameter becomes 
a genuine conserved order parameter dynamics. One can in fact observe 
multiscaling, if the volume of the systems is large enough. 
The existence of inhomogeneous structures is mirrored in the 
presence of the peak
in the structure factor at finite wavelength and the phenomenon is similar
to the Ostwald ripening. The undercooling initially present is not
sufficient to promote the transformation 
of all the liquid into the solid state, and some
drops of either phase remain trapped into the other.

To summarize the results we have established the following 
rules governing the evolution:

1) The field $U$ is constant in time. 

2) The order parameter $m(t)$ as $t\to \infty$ tends to the asymptotic
value $m=2 Q$, if $Q$ falls in the range $[-1/2,1/2]$. This fact
in turn implies that the spatial average value of $u$ over the system 
vanishes as $t \to 0$ , 
i.e. the system reaches two phase coexistence asymptotically.

3) In the above range of $Q$, the correlation function is large
and centered at finite values of the wavevector $k$.

4) If $|Q|$ exceeds the threshold value $1/2$ the
system evolves towards a spatially uniform state with $m\sim -1$ and
vanishing correlations and $m$
is no longer equal to $2 Q$. In this case $u$ reaches an equilibrium value
which is non zero and the system is out of two phase coexistence. 

5) If $|Q|>1/2$, i.e., larger  values of the undercooling
cause the melt to crystallize completely and thus correlations
are asymptotically suppressed as the system reaches an homogeneous state.

The above features are interesting because
mimic the behavior of the more realistic Phase field model with local
constraint. 

We, finally, remark that perhaps, the most serious flaw of the
model, is that it suffers from the same problem as the spherical model. 
In contrast with the local constrained Phase field model, 
which displays a region of metastability of the liquid phase in the
$\phi$-$T$ plane , between the
coexistence line and the spinodal line \cite{Binder}, the globally
constrained model is always unstable inside the two phase coexistence line.
As a consequence,
no nucleation barrier needs to be overcome in the transformation 
from liquid to solid. In the initial state long-wavelength fluctuations 
grow and the system becomes unstable.
The nucleation barrier being proportional to the surface tension associated 
with the creation of a kink in the scalar model, whereas in the model
with global couplings the energy gaps between the ordered phase
and the instanton solutions, i.e. the uniform solutions of the
equation vanishes in the infinite volume limit.
Thus the mechanism described is non-Arrhenius like.

This is also reflected in the absence of true
phase separation, since the width of a domain wall
diverges in the limit of a  vanishing pinning field, $h$,
as $h^{-1/2}$ \cite{Abraham}.

\begin{figure}
\caption{Schematic behaviour of 
         $\partial m / \partial t$ as a function of $m$,
         from eq. (\protect\ref{eq:mm}).
        }
\label{fig:mdot}
\end{figure}

\begin{figure}
\caption{Typical behavious of the order parameter $m$ as function of time
         for the cases:
         (a) $Q>1/2$; 
         (b) $0<Q<1/2$; 
         (c) $-1/2<Q<0$; 
         (d) $Q<-1/2$.
         In all runs shown here we started with $m(t=0) = 0.54$, while the other 
         parameter were:
         (a)  $\overline{u}(t=0) =  0.4$, $Q =  0.67$, $m(\infty) \not= 2Q$;
         (b)  $\overline{u}(t=0) = -0.2$, $Q =  0.07$, $m(\infty) = 2Q$;
         (c)  $\overline{u}(t=0) = -0.4$, $Q = -0.13$, $m(\infty) = 2Q$;
         (d)  $\overline{u}(t=0) = -1.0$, $Q = -0.73$, $m(\infty) \not= 2Q$.
        }
\label{fig:figG1}
\end{figure}

\begin{figure}
\caption{The order 
         parameter $m$ as a function of time for $Q=-0.3077$ and different
         values of $\lambda$, 
         from rigth to left $\lambda=$ $128$, $170$, $200$, $300$, $500$.
         The plateau increases as $\lambda$ decreases.
         At the initial time the parameters are $m=0.385$ and $u=-0.5$.
	 The dashed line denotes the value $2Q$.
        }
\label{fig:figA1}
\end{figure}

\begin{figure}
\caption{The cross-over time $t_c$ as function of $\lambda$ for the
         curves of Fig. \protect\ref{fig:figA1}. The dashed line is the
         scaling $1/\lambda$. The cross-over time is defined as
         $\min_{t}\,m(t)=2Q$.
        }
\label{fig:figA2}
\end{figure}

\begin{figure}
\caption{Snapshot of the $\phi$ field for the global constraint case and 
         $-1/2< Q < 0$. On the axis we report the lattice index.
        }
\label{fig:figG2}
\end{figure}

\begin{figure}
\caption{Snapshot of the $\phi$ field for the local constraint case and
         $-1/2 <Q < 0$. On the axis we report the lattice index.
        }
\label{fig:figG3}
\end{figure}

\begin{figure}
\caption{Circularly averaged $\phi\phi$ correlation function as function of
         the lattice index for the 
         global constraint case obtained numerically: 
         circles represent $|Q|<1/2$; diamonds refer to $|Q|>1/2$.
        }
\label{fig:figG4}
\end{figure}

\begin{figure}
\caption{Circularly averaged $\phi\phi$ correlation function as function 
         of the lattice index for the 
         local constraint case obtained numerically: circles refer to $|Q|<1/2$
         whereas  diamonds $|Q|>1/2$.
        }
\label{fig:figG5}
\end{figure}


\begin{references}
\bibitem{sferico}
        The present spherical model defined by 
        eq.(\protect\ref{eq:hh}) is different from the $O(n)$ spherical
        model discussed previous papers \protect\cite{UMBM}. In that case 
        both the order parameter and the $U$ field were vector fields
        coupled bilinearly. However, the two models in the infinite
        volume limit and in the limit of infinite components generate
        identical equations for the average fields and their correlations,
        although the physical interpretation remains different for the 
        two models.
\bibitem{precisazione}
         Notice that the values of the coupling parameters 
         of the $U^2$ term and $U\phi$, in eq (\protect\ref{eq:hh}) 
         have not been chosen independently as assumed in 
         ref. \cite{UMBM}, but their ratio was fixed as to guarantee 
         the correct energy balance as
         explained in Sections 
         \protect\ref{sec:PhField}-\protect\ref{sec:ThermP}.
\end{references}
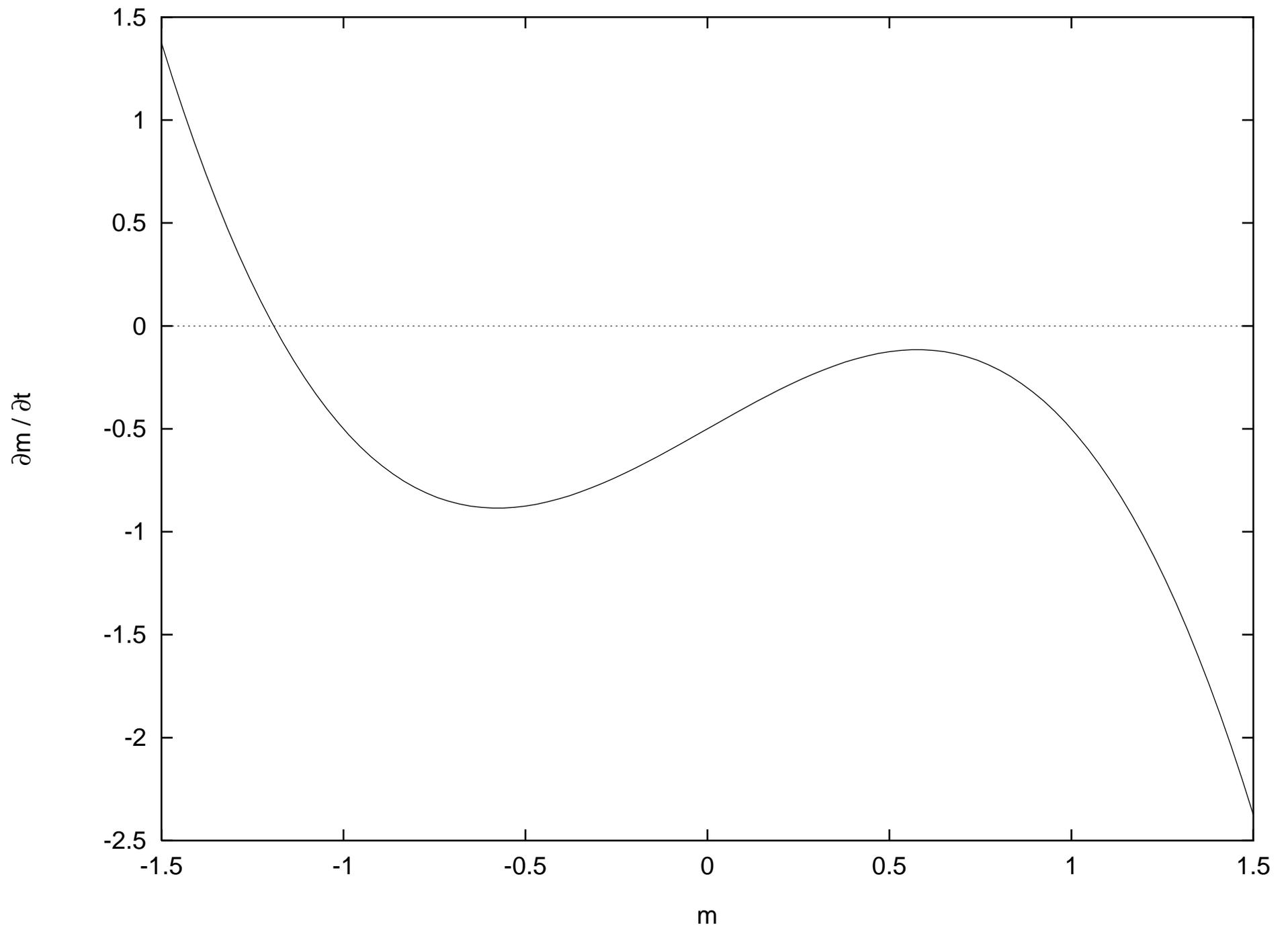
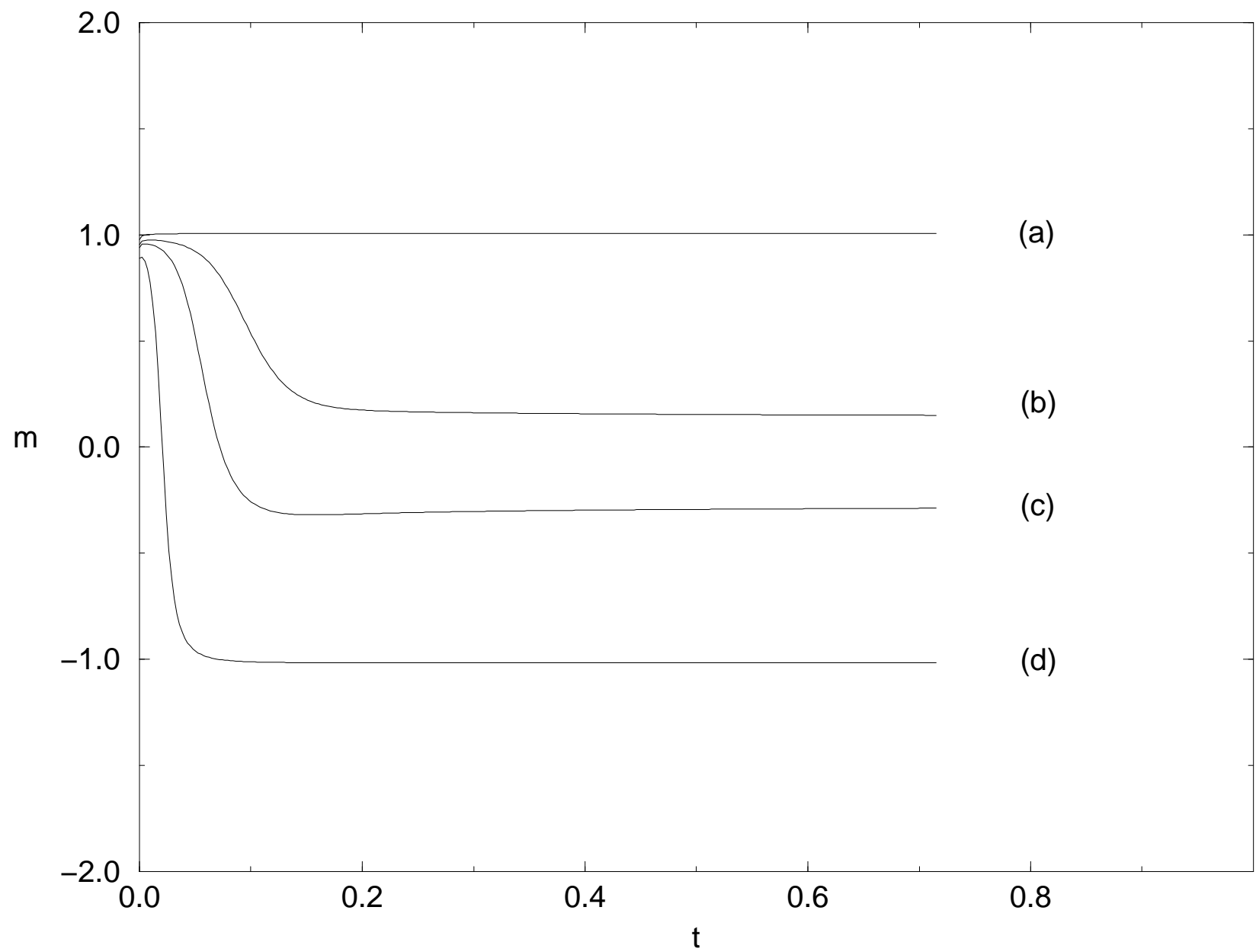
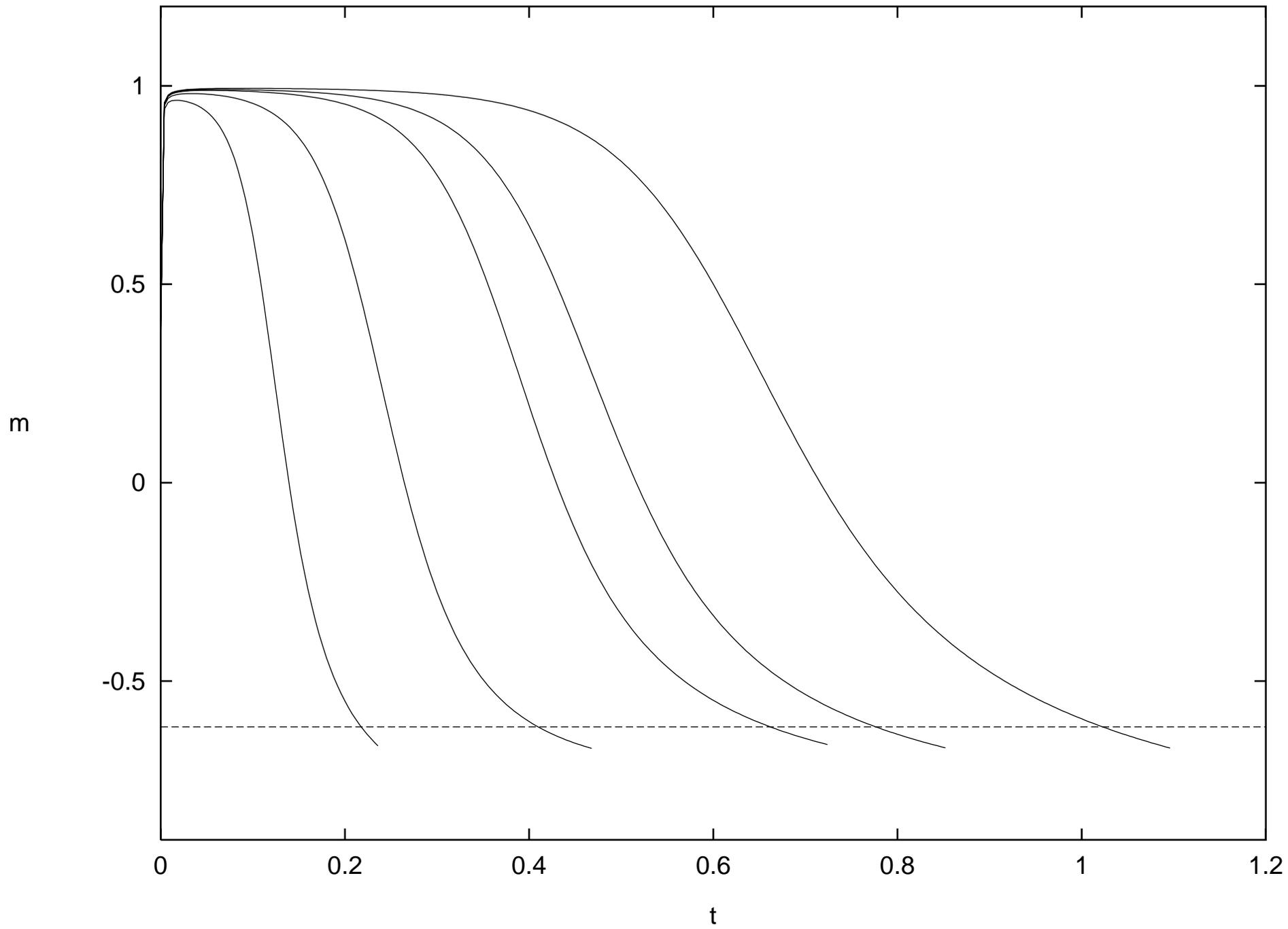
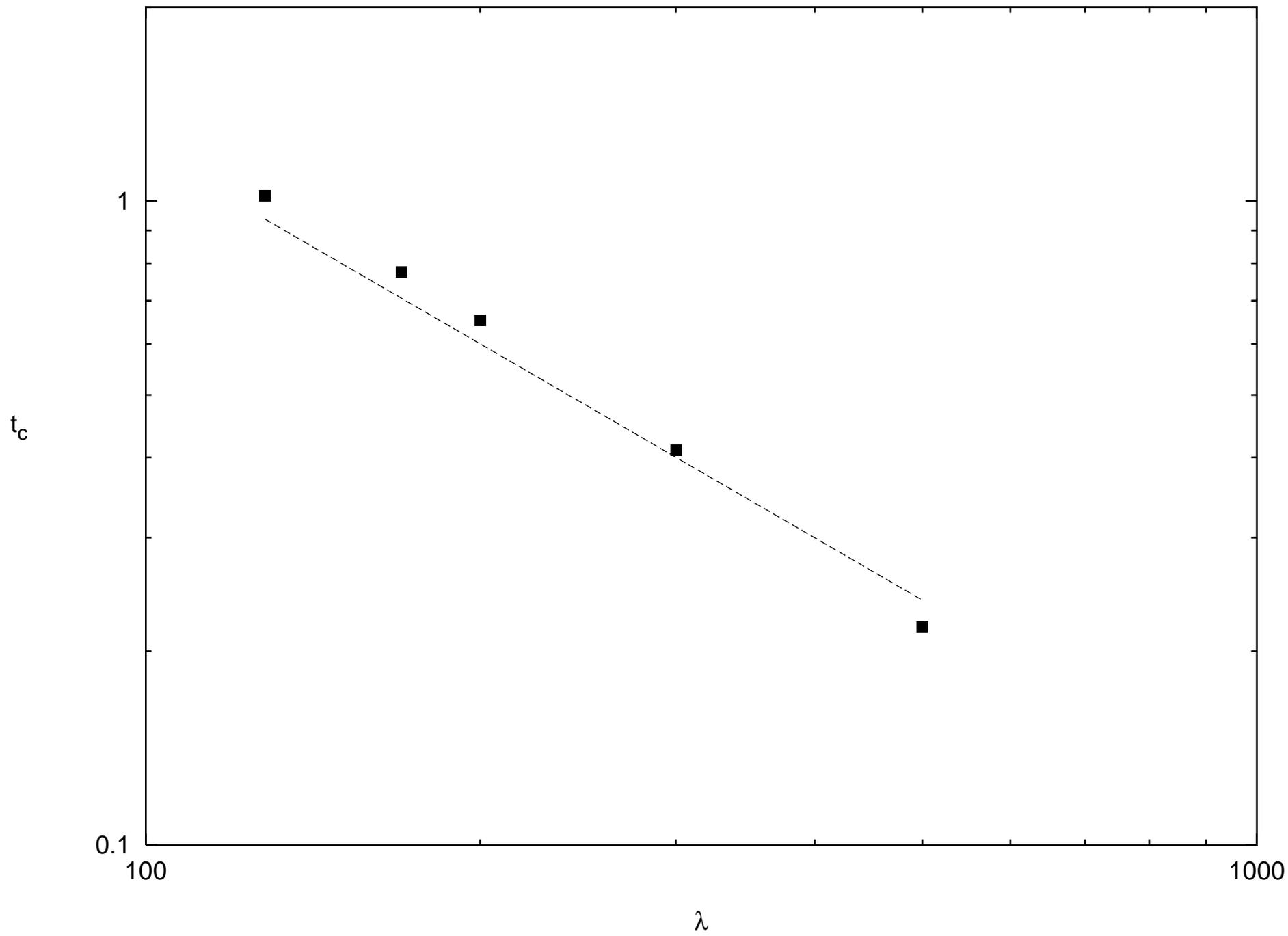
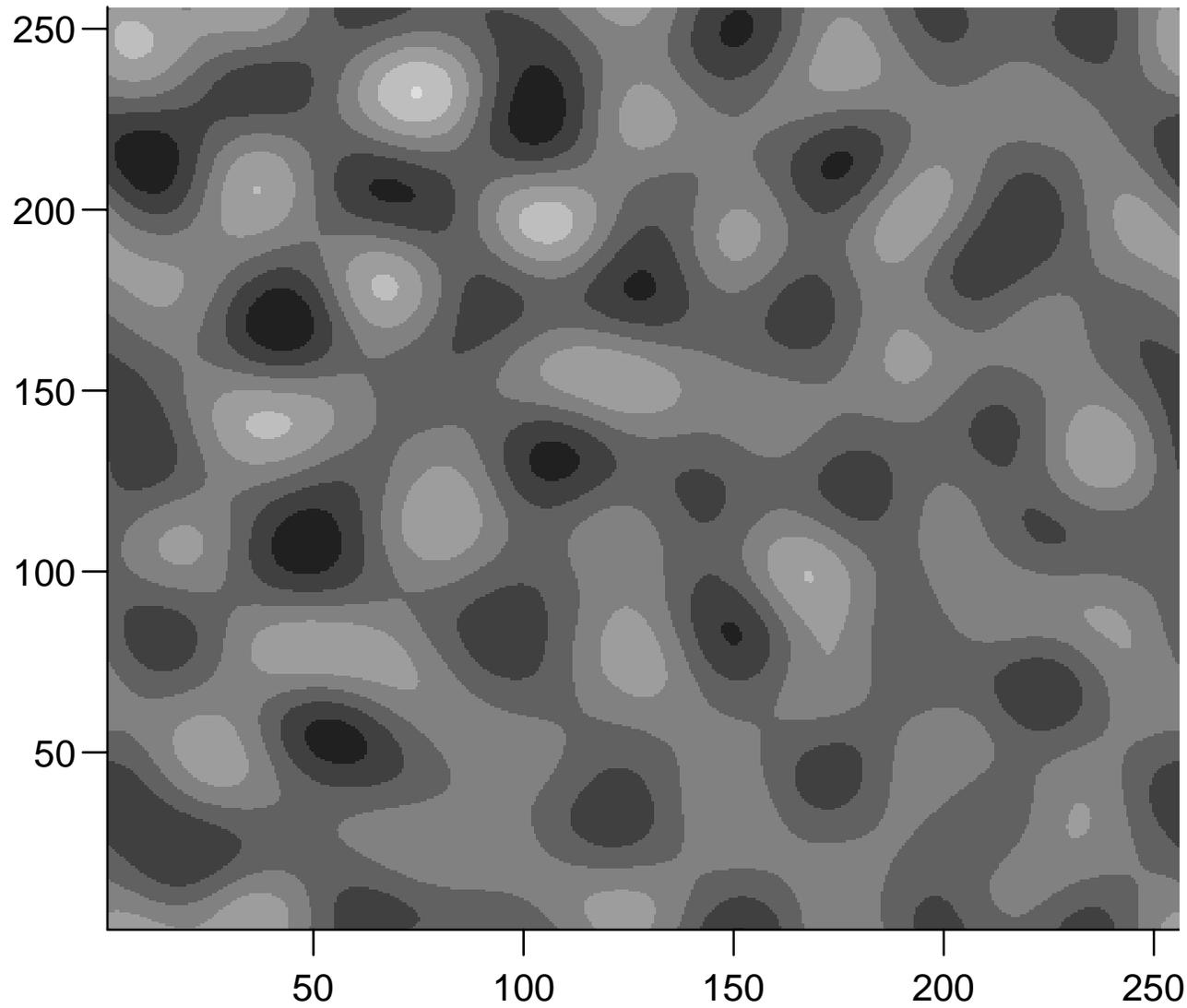
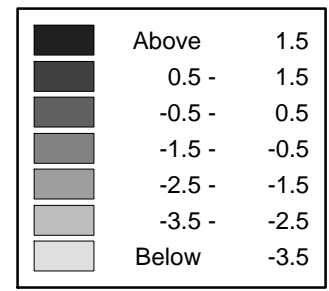
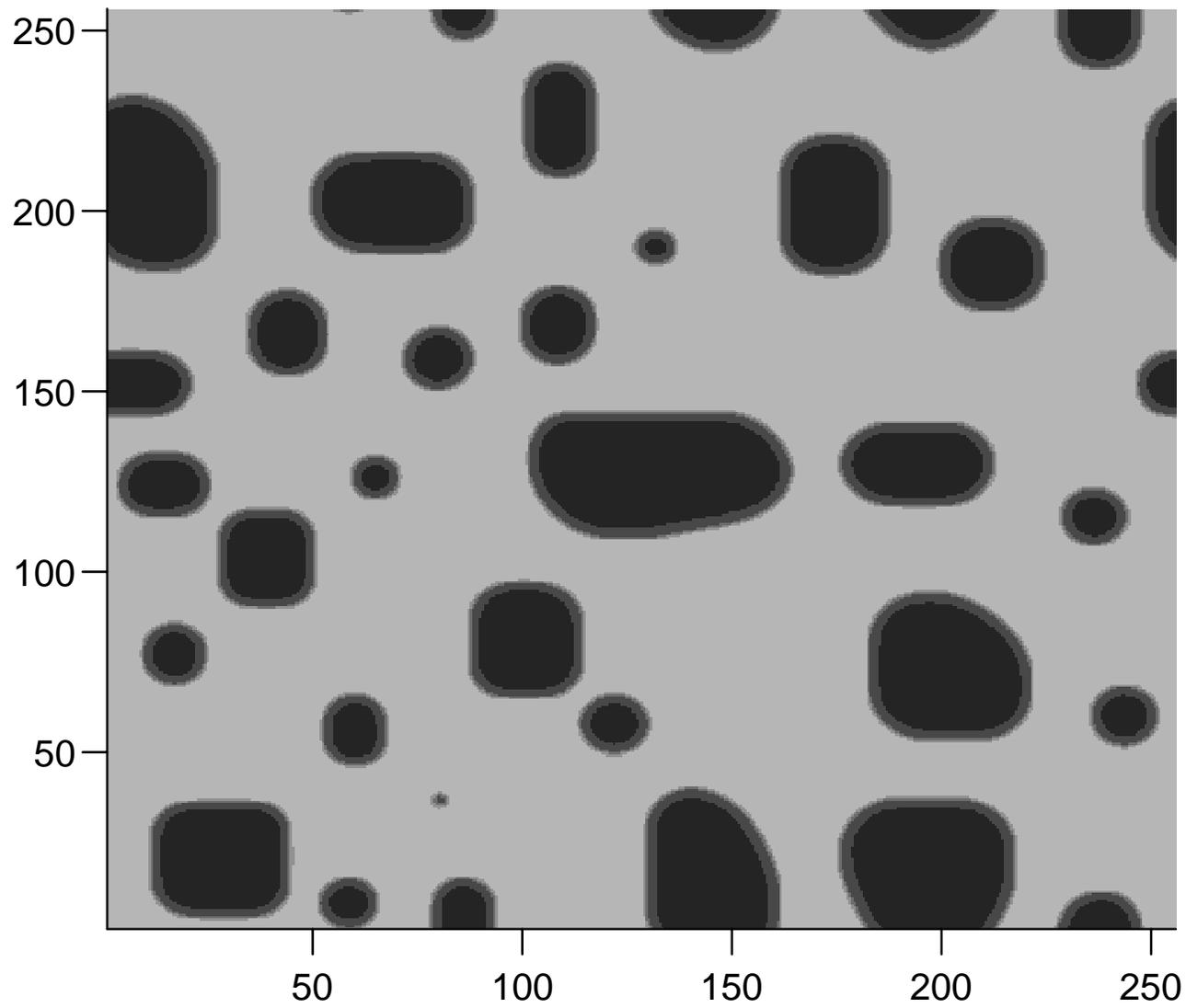
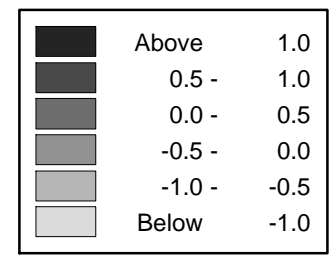
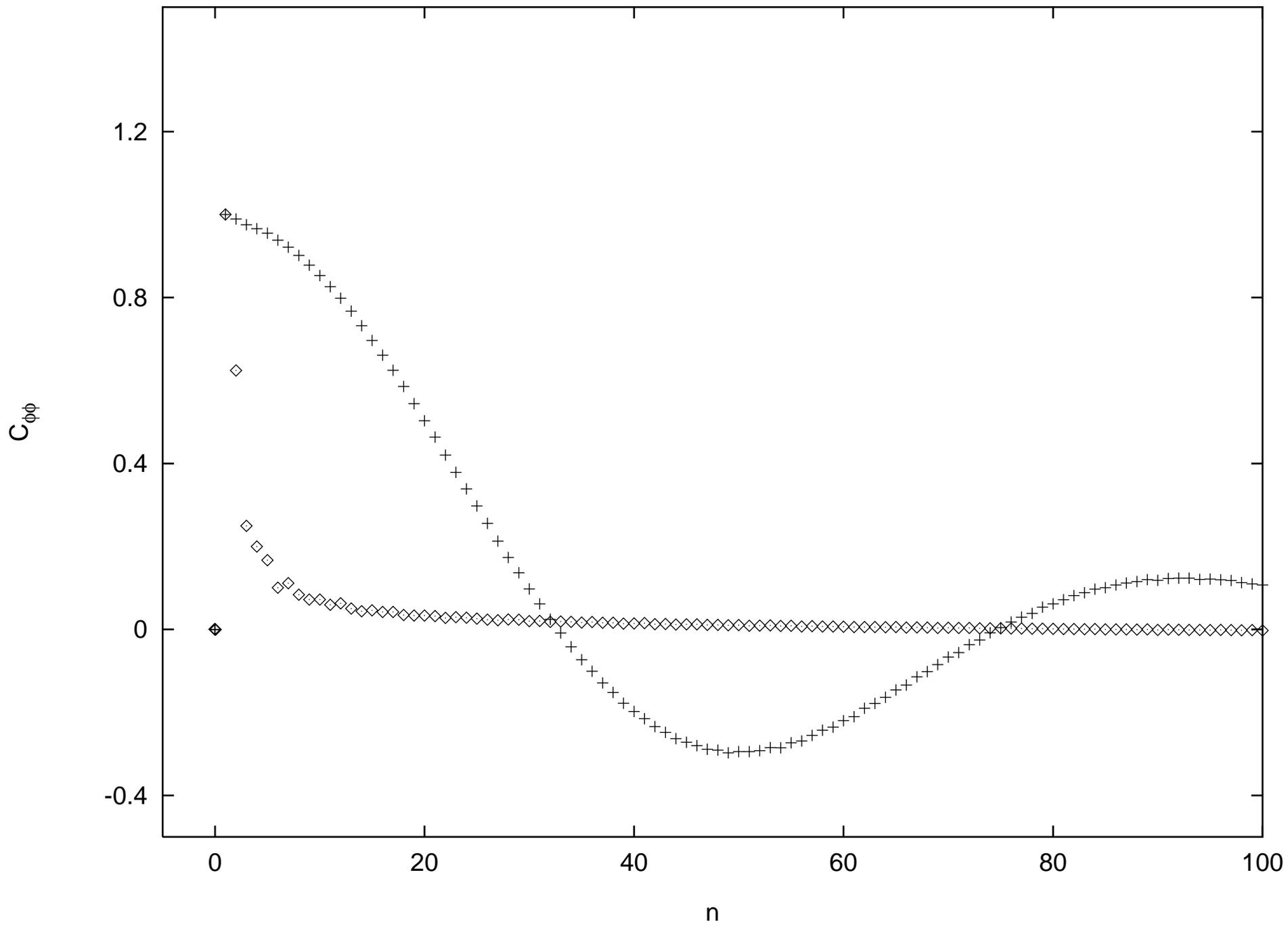
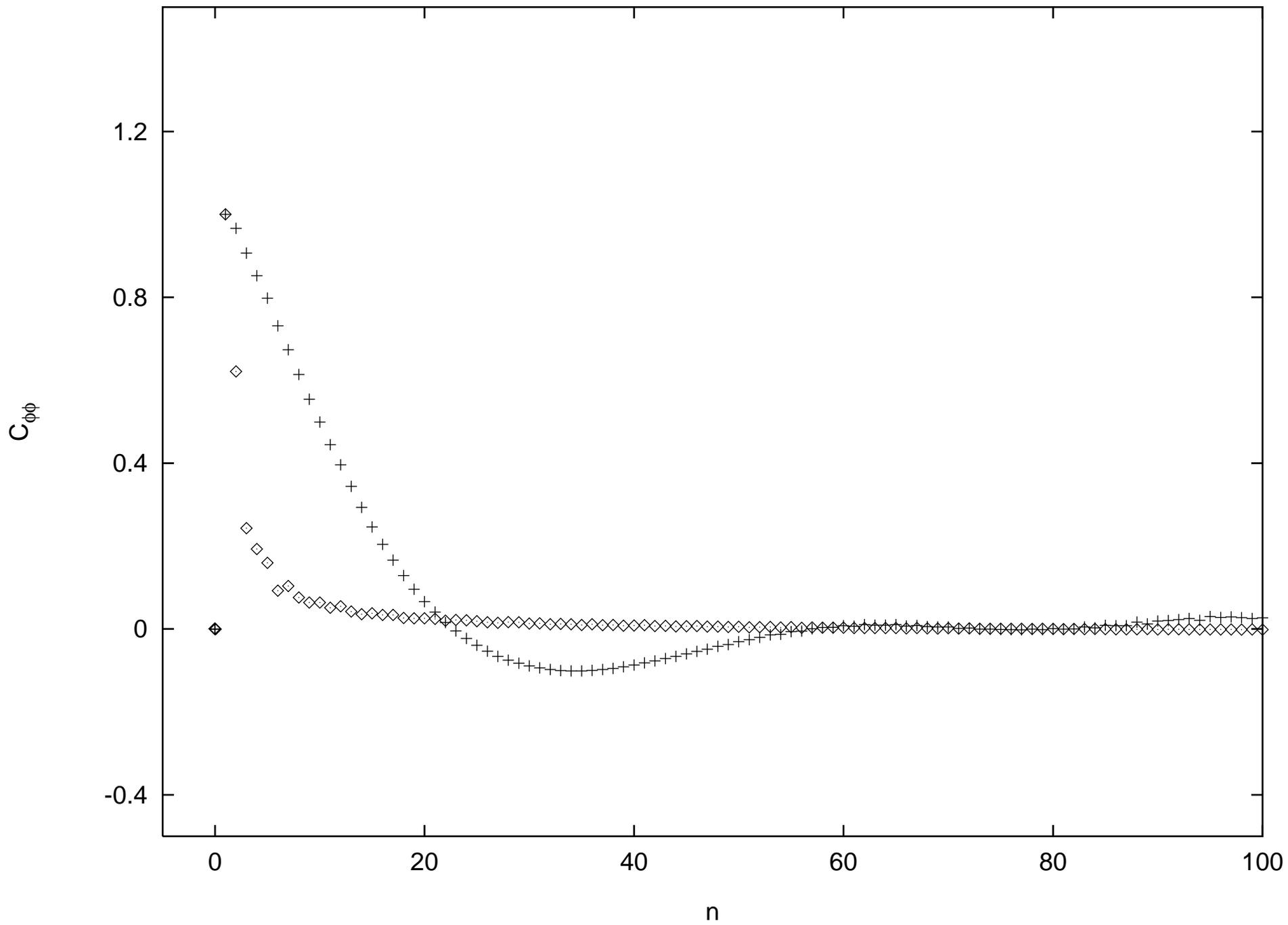
\end{document}